\documentclass[conference, 10pt]{IEEEtran}
\IEEEoverridecommandlockouts
\usepackage[T1]{fontenc}
\usepackage[cmintegrals]{newtxmath}
\usepackage{bm}
\usepackage[numbers, sort&compress]{natbib}
\usepackage{amsmath,amsfonts}
\usepackage{tabularx, booktabs, multirow}
\usepackage{siunitx}
\newcolumntype{Y}{>{\centering\arraybackslash}X}
\newcolumntype{Z}{>{\raggedleft\arraybackslash}X}
\newcolumntype{L}{>{\raggedright\arraybackslash}X}
\usepackage{graphicx}
\usepackage[caption=false]{subfig}
\usepackage[export]{adjustbox}
\usepackage{pgfplots}
\usepgfplotslibrary{groupplots}
\pgfplotsset{compat=1.18}
\usepackage{tikz}
\usetikzlibrary{arrows.meta, positioning, shapes.geometric, shapes, calc}
\newsavebox{\hsboxempty}
\newsavebox{\hsboxhalf}
\newsavebox{\hsboxfull}
\AtBeginDocument{%
  \sbox{\hsboxempty}{\tikz[baseline=-0.75ex]{\draw[thin](0,0)circle(.38em);}}%
  \sbox{\hsboxhalf}{\tikz[baseline=-0.75ex]{%
    \draw[thin](0,0)circle(.38em);%
    \fill(0,0)--(90:.38em)arc(90:270:.38em)--cycle;}}%
  \sbox{\hsboxfull}{\tikz[baseline=-0.75ex]{\fill(0,0)circle(.38em);}}%
}
\newcommand{\emptycirc}{\usebox{\hsboxempty}}
\newcommand{\halfcirc}{\usebox{\hsboxhalf}}
\newcommand{\fullcirc}{\usebox{\hsboxfull}}
\usepackage{xcolor}
\definecolor{hsurot}   {cmyk}{.00, 1, .59, .26}
\definecolor{hsugrau}  {cmyk}{.38, .37, .39, .15}
\definecolor{hsugelb}  {cmyk}{0, .16, .80, 0}
\definecolor{hsublau}  {cmyk}{1, .40, 0, .82}
\definecolor{hsuturkis}{cmyk}{1, .14, .60, .49}
\definecolor{hsugruen} {cmyk}{.16, .16, .91, .28}
\definecolor{hsuorange}{cmyk}{.01, .87, .77, .13}
\usepackage[printonlyused, nohyperlinks]{acronym}
\acrodef{lem}[LEM]{local electricity market}
\acrodef{pv}[PV]{photovoltaic}
\acrodef{bess}[BESS]{battery energy storage system}
\acrodef{ev}[EV]{electric vehicle}
\acrodef{soc}[SOC]{state of charge}
\acrodef{kpi}[KPI]{key performance indicator}
\acrodef{hp}[HP]{heat pump}
\acrodef{jade}[JADE]{Java Agent Development Framework}
\acrodef{ad}[AD]{autarky degree}
\acrodef{scr}[SCR]{self-consumption rate}
\acrodef{fgp}[FGP]{financial gain through participation}
\acrodef{mr}[MR]{match rate}
\usepackage{todonotes}
\usepackage{enumitem}
\usepackage[english]{babel}
\usepackage{hyperref}
\hypersetup{colorlinks=true, unicode=true,
            linkcolor=[rgb]{0.10,0.05,0.67},
            citecolor=[rgb]{0.10,0.05,0.67},
            filecolor=[rgb]{0.10,0.05,0.67},
            urlcolor=[rgb]{0.10,0.05,0.67}}
\addto\extrasenglish{}

\addto\extrasenglish{}
\addto\extrasenglish{}
\addto\extrasenglish{}
\addto\extrasenglish{}
\begin{document}

\title{Market Strategy Evaluation for Prosumers in Local Electricity Markets%
\thanks{This research in the project OptiFlex was funded by dtec.bw
(Digitalization and Technology Research Center of the Bundeswehr). dtec.bw is funded
by the European Union (NextGenerationEU). This work was also supported by the
Federal Ministry for Economic Affairs and Climate Action and the Projektträger
Jülich GmbH within the project InterPhaSe (FKZ 03EI6139A).}}

\author{%
\IEEEauthorblockN{Lukas P.~Wagner, Raoul Bisson, and Felix Gehlhoff}
\IEEEauthorblockA{Institute of Automation Technology, \\ Helmut Schmidt University, Hamburg, Germany\\  \{lukas.wagner, raoul.bisson, felix.gehlhoff\}@hsu.hamburg}
}

\maketitle

\begin{abstract}
Prosumers equipped with distributed generation and flexible loads form
autonomous cyber-physical energy systems that control local resources and
participate in local energy markets with minimal human intervention. This work
develops and evaluates an agent-based simulation platform in which agents,
representing prosumer households with photovoltaic systems, battery storage
systems, electric vehicles, and heat pumps, participate in a uniform-price
double-sided call auction. The effect of individual bidding strategies on
community-level efficiency and prosumer-level financial outcomes is incompletely
understood, particularly when prosumers with heterogeneous portfolios interact in
one market. Four market
strategies of increasing complexity are compared: a zero-intelligence
constrained baseline, a boundary-price strategy, an extended storage cascade, and
a market-adaptive pricing strategy. The simulation is conducted on a community of
33 prosumers at 15-minute resolution, spanning summer, winter, and spring to characterize seasonal variation. Results show that rule-based resource
control substantially reduces community energy expenditure: the extended storage
cascade achieves a total cost of 39.06\,EUR compared to 62.38\,EUR under the
zero-intelligence baseline, a reduction of 37.4\,\%. The market-adaptive strategy
yields the highest aggregate community financial gain through local energy market participation
(14.40\,EUR vs.\ 10.28\,EUR for the baseline, a gain of 40.1\,\%) under summer conditions.
Strategy effectiveness depends on both portfolio composition and seasonal supply conditions, requiring joint evaluation of resource control and pricing decisions.
\end{abstract}


\section{Introduction}
\label{sec:intro}

The decarbonization of residential energy supply drives rapid expansion of
distributed generation and flexible consumption at the low-voltage level.
Rooftop \ac{pv} systems,  \acp{bess}, \acp{ev}, and \acp{hp}
transform a growing number of households into prosumers that independently
control local energy resources, respond to market signals, and exchange energy with
minimal human intervention. 
 \Acp{bess} and \acp{ev} are key enablers of \ac{pv}
self-consumption \cite{LUTHANDER2015_Photovoltaicselfconsumptioninbuildings},
as they introduce additional temporal flexibility for shifting locally generated
energy.

In Germany alone,
more than 3.7 million residential \ac{pv} installations were registered by
2023, with shares of electrified heating and mobility rising in parallel \cite{Kil_IntegratingPeer-to-PeerEnergyTradingandFlexibility_2023}.
Without incentive-compatible coordination mechanisms, surplus generation is
curtailed or exported at low feed-in tariffs while neighbors simultaneously
import energy at the substantially higher retail tariff \cite{Kil_IntegratingPeer-to-PeerEnergyTradingandFlexibility_2023}. \Acp{lem} address this misalignment by enabling direct energy exchange within a geographically
bounded community, typically a low-voltage feeder or neighborhood, at negotiated
prices between prosumers \cite{Mengelkamp2019_TracingLocalEnergyMarkets}. Survey work by
\citet{Mengelkamp2019_TracingLocalEnergyMarkets} identifies double-sided auctions
as the dominant clearing mechanism and agent-based simulation as the primary
methodological tool for \ac{lem} analysis: in double-sided auctions, software agents are well suited for
representing prosumers in this context. \citet{Reinpold2025_SystematiccomparisonofsoftwareagentsandDigitalTwins}
identify autonomy, external resource control, and sociability as defining agent properties, which correspond
directly to a prosumer's need to independently control local resources while
interacting with an \ac{lem}.

This work is embedded in the InterPhaSe project,
which investigates an integrated platform combining an \ac{lem} with market-based
distribution grid congestion management. Prior contributions addressed
decentralized energy dispatch and scalable agent-based control
\cite{Kil_ADecentralizedOptimizationApproachForScalable_2025} as well as
identity-secured market access \cite{Kil_IntegratingPeer-to-PeerEnergyTradingandFlexibility_2023}.
The present work addresses the prosumer level: given a specific resource
portfolio, which market strategy should a prosumer adopt to maximize its financial
benefit while contributing to community efficiency?

This question involves two intertwined sub-problems. The resource control problem
asks how flexible resources should be dispatched within each interval to minimize
net energy cost. The bid pricing problem asks at what price the agent should
submit its sell offer or buy request to the auction. These decisions interact: local storage
reduces the quantity available for market sale, thereby affecting price discovery,
while the achieved clearing price determines the economic value of stored energy
and thus the optimal dispatch rule. Compounding this interaction is portfolio
heterogeneity. A prosumer combining a \ac{pv} system with a \ac{bess}
and an \ac{ev} can absorb surplus along two independent storage pathways and has
multiple options for intra-day arbitrage, whereas a \ac{pv}-only prosumer is limited
to direct market sales, and a pure consumer acts exclusively as a buyer.
This motivates a comparative evaluation that spans heterogeneous portfolio
classes rather than assuming a homogeneous prosumer population.

Evaluating such prosumers requires simulation environments that faithfully
reproduce both resource dynamics and the market interaction layer. Existing simulation studies compare strategies for homogeneous
populations or evaluate market mechanisms without distinguishing the portfolio
dimension. A joint simulation-based evaluation of resource control and bid pricing
strategies across heterogeneous portfolio classes, covering both community-level
cost and per-portfolio financial gain, is absent from the literature. 

Three research questions (RQs) guide this work:
\begin{enumerate}[label={RQ\arabic*.}]
  \item To what extent do resource control and bid pricing rules each contribute
    to reducing community energy cost?
  \item How does an extended storage cascade, a sequential dispatch policy
    prioritizing \ac{bess} and then \ac{ev} storage (defined in \autoref{sec:strategies}), incorporating the \ac{ev} battery
    affect autarky across portfolio classes?
  \item How does market-adaptive pricing redistribute financial gains, and how
    sensitive is this effect to community composition?
\end{enumerate}

The contributions of this work are:
\begin{enumerate}[label=(\roman*)]
  \item An agent-based \ac{lem} simulation platform implemented with \ac{jade}
    \cite{Bel_DevelopingMultiAgentSystemswithJADE_2007},
    featuring configurable, portfolio-specific resource models and bidding strategies.
  \item A strategy framework that independently varies resource
    control and bid pricing, enabling targeted attribution of each dimension's
    contribution to community cost and financial gain.
  \item A multi-\ac{kpi} evaluation across five heterogeneous portfolio classes
    and three seasonal scenarios, quantifying community cost, autarky, market
    participation, and per-portfolio financial outcomes.
\end{enumerate}

The remainder of this paper is organized as follows. \autoref{sec:related}
surveys related work and identifies the research gap. \autoref{sec:system}
describes the agent architecture, resource models, and market mechanism.
\autoref{sec:strategies} defines the four strategies. \autoref{sec:eval}
presents the simulation setup and evaluation results. \autoref{sec:conclusion} summarizes findings
and outlines future work.

\section{Related Work}
\label{sec:related}

This section reviews related work on agent-based \ac{lem} simulation and
prosumer bidding strategies, structured around four requirements (Rs) motivated
by the joint resource-control and bid-pricing problem introduced in
\autoref{sec:intro}. These requirements jointly enable a structured evaluation
of market strategies, including the attribution of system-level effects to
resource control and pricing decisions as well as the analysis of
portfolio-dependent outcomes:
\begin{enumerate}[label={R\arabic*.}]
  \item Agent-based prosumer modeling with portfolio-specific resource configurations,
    because strategy effectiveness depends on asset configuration
    \cite{Mengelkamp2018_IntelligentStrategiesforResidentialCustomers}.
  \item A uniform-price double-sided call auction as clearing mechanism, as this
    is the dominant design in the \ac{lem} literature and determines the
    incentive structure under study
    \cite{Mengelkamp2019_TracingLocalEnergyMarkets}.
  \item Multiple strategies with independently configurable and analyzable resource control and
    bid pricing dimensions, to isolate the contribution of each to observed
    outcomes \cite{SCHOLZEL2024Comparativeanalysisofbiddingstrategies}.
  \item Quantitative \ac{kpi} evaluation covering economic efficiency, autarky,
    and local market participation across multiple seasonal scenarios, enabling
    objective and reproducible strategy comparison
    \cite{Kil_IntegratingPeer-to-PeerEnergyTradingandFlexibility_2023}.
\end{enumerate}

\citet{Kil_IntegratingPeer-to-PeerEnergyTradingandFlexibility_2023} present an
agent-based platform for peer-to-peer energy trading combined with congestion
management, evaluating prosumer agents with self-consumption optimization and
forecast-based pricing across a 16-node distribution grid (R1). A single
strategy variant is implemented without a central auction mechanism, so neither
the clearing design (R2) nor a comparison of independently configured strategies
(R3) is addressed. Quantitative metrics are reported but not disaggregated by
portfolio class or season (R4 partial).

\citet{Kil_ADecentralizedOptimizationApproachForScalable_2025} develop a
decentralized two-agent architecture per household for simultaneous energy
dispatch and congestion management on the IEEE 33-bus network (R1). No central
auction or strategy comparison is performed (R2, R3 absent), and evaluation
covers a single scenario without portfolio-class or seasonal breakdown
(R4 partial).

\citet{Mengelkamp2018_IntelligentStrategiesforResidentialCustomers} simulate 100
households in a merit-order \ac{lem} and compare reinforcement learning strategies
with rule-based variants, reporting outcomes separately for consumer and prosumer
types (R2, R4). All households rely on a shared community storage rather than
individual storage configurations, limiting portfolio heterogeneity (R1 partial).
Furthermore, resource control and bid pricing are not independently varied,
preventing attribution of observed effects to either dimension (R3 partial).

\citet{ELBAZ2019Integrationofenergymarketsinmicrogrids} design resource-type-specific
bidding strategies for a uniform-price double-sided auction covering \ac{pv},
\ac{hp}, \ac{ev}, and \ac{bess}, reporting per-resource-class cost
reductions (R2, R4). Since all households carry the same full resource set,
portfolio-class effects cannot be isolated (R1 partial), and resource control and
bid pricing are not independently configurable (R3 partial).

\citet{SCHOLZEL2024Comparativeanalysisofbiddingstrategies} compare
zero-intelligence, reinforcement learning, and device-oriented strategies in a
uniform-price auction with quantitative evaluation (R2, R4). All households
are identically equipped, precluding portfolio differentiation (R1 absent), and
resource dispatch and bid pricing are not varied independently (R3 partial).

\citet{Bose2021_ReinforcementLearningLocalEnergyMarkets} and
\citet{Okwuibe2022_IntelligentBiddingStrategiesforProsumers} apply reinforcement
learning in \acp{lem} and report autarky and cost \acp{kpi} (R2). Agent types
are distinguished as prosumer or consumer rather than by specific resource
portfolios (R1 partial), and both approaches optimize a single combined policy
without separating resource control from bid pricing, limiting interpretability
of strategy effects (R3 partial). Local market participation metrics are not the
primary evaluation target (R4 partial).

\begin{table}[h!]
\caption{Coverage of key requirements in related work.}
\label{tab:relwork}
\footnotesize
\setlength{\tabcolsep}{4pt}
\begin{tabularx}{\columnwidth}{Lcccc}
\toprule
Reference & R1 & R2 & R3 & R4 \\
\midrule
\citet{Kil_IntegratingPeer-to-PeerEnergyTradingandFlexibility_2023} & \fullcirc & \emptycirc & \emptycirc & \halfcirc \\
\citet{Kil_ADecentralizedOptimizationApproachForScalable_2025} & \fullcirc & \emptycirc & \emptycirc & \halfcirc \\
\citet{Mengelkamp2018_IntelligentStrategiesforResidentialCustomers} & \halfcirc & \fullcirc & \halfcirc & \halfcirc \\
\citet{ELBAZ2019Integrationofenergymarketsinmicrogrids} & \halfcirc & \fullcirc & \halfcirc & \halfcirc \\
\citet{SCHOLZEL2024Comparativeanalysisofbiddingstrategies} & \emptycirc & \fullcirc & \halfcirc & \fullcirc \\
\citet{Bose2021_ReinforcementLearningLocalEnergyMarkets} & \halfcirc & \fullcirc & \halfcirc & \halfcirc \\
\citet{Okwuibe2022_IntelligentBiddingStrategiesforProsumers} & \halfcirc & \fullcirc & \halfcirc & \halfcirc \\
\bottomrule
\end{tabularx}
{\footnotesize\fullcirc{} satisfied\enspace\halfcirc{} partial\enspace\emptycirc{} not addressed}
\end{table}

As shown in \autoref{tab:relwork}, no existing work simultaneously satisfies
all four requirements. Existing studies either evaluate market strategies
under homogeneous agent assumptions or implicitly couple resource control and
bid pricing within a single policy. As a result, the contribution of each
dimension to system-level performance cannot be isolated.

In particular, no study combines a uniform-price call auction (R2) with
independently configurable resource control and pricing strategies (R3) across a
heterogeneous prosumer population (R1). No study reports \ac{kpi} results
disaggregated by portfolio class (R4), leaving the portfolio-dependent
distribution of gains and the efficiency-benefit trade-off uncharacterized.

\section{System Model, Market Mechanism, and Performance Metrics}
\label{sec:system}

This section describes the simulation platform: the multi-agent architecture
and interaction protocol (\autoref{subsec:arch}), the resource models
of each prosumer (\autoref{subsec:devices}), the uniform-price call auction
used for market clearing (\autoref{subsec:market}), and the \acp{kpi} used
for strategy evaluation (\autoref{subsec:kpis}).

\subsection{Agent Architecture}
\label{subsec:arch}

\begin{figure}[t]
\centering
\resizebox{\columnwidth}{!}{%
\includegraphics{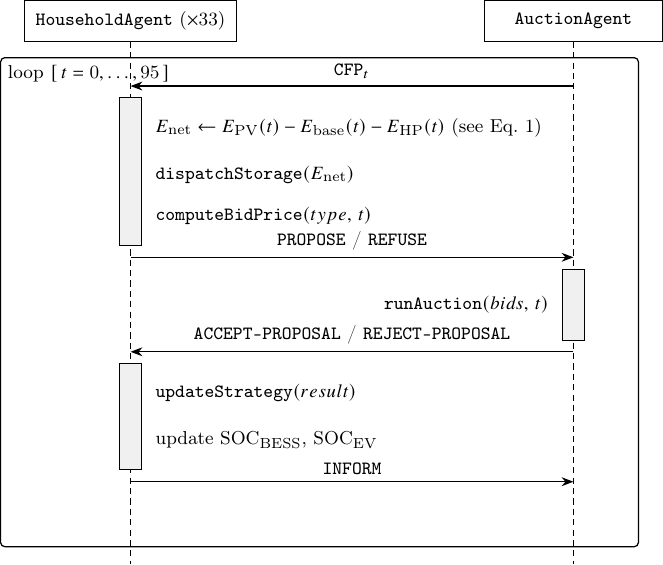}%
}
\caption{FIPA Contract Net interaction protocol for one slot $t$.
Activation boxes indicate local processing steps.}
\label{fig:protocol}
\end{figure}

The simulation is implemented as a multi-agent system using \ac{jade}
\cite{Bel_DevelopingMultiAgentSystemswithJADE_2007}, a FIPA-compliant platform
for asynchronous agent communication language messaging. The platform instantiates
two agent types: 33 \texttt{HouseholdAgent}s, each representing one prosumer
household, and a single \texttt{AuctionAgent} that coordinates the market.
The per-slot interaction follows the FIPA Contract Net Interaction Protocol
\cite{Smith1980_ContractNetProtocol} as a communication and coordination layer,
while the allocation decision is realized through a centralized uniform-price
auction (see \autoref{subsec:market}). \autoref{fig:protocol} shows the message exchange
for one slot.

Each of the $T = 96$ slots of $\Delta t = 15\,\text{min}$ follows a synchronous
round opened by the \texttt{AuctionAgent} broadcasting a call for proposals
(\texttt{CFP}). Each
\texttt{HouseholdAgent} responds by computing its net energy balance
$E_\text{net}$ from local generation and consumption (\autoref{eq:balance}),
then dispatching \ac{bess} and \ac{ev} storage according to its
assigned strategy. If the residual after storage dispatch satisfies
$|E_\text{net}| \geq \varepsilon$ ($\varepsilon = 0.001\,\text{kWh}$, the
minimum tradeable quantity), the agent determines a bid price and submits a
proposal (\texttt{PROPOSE}) carrying bid type (sell or buy), quantity, and price. If
storage dispatch absorbs the full surplus or covers the full deficit within the
slot, $|E_\text{net}|$ falls below $\varepsilon$ and the agent sends
a refusal (\texttt{REFUSE}). This is a per-slot condition and does not imply autarky: the
same agent may still import or export in other slots, or fail to find a match
and fall back to the grid tariff.

After collecting all proposals, the \texttt{AuctionAgent} evaluates them
jointly via merit-order clearing (see \autoref{subsec:market}) and notifies every agent: matched prosumers receive
\texttt{ACCEPT-PROPOSAL} with clearing price and traded quantity; unmatched prosumers receive
\texttt{REJECT-PROPOSAL} together with the slot clearing price, which
can be used for strategy adaptation in the following slot. Each
\texttt{HouseholdAgent} then adjusts its internal strategy state, updates
the \ac{soc} of \ac{bess} ($\text{SOC}_\text{BESS}$) and \ac{ev} ($\text{SOC}_\text{EV}$), and sends \texttt{INFORM} to the
\texttt{AuctionAgent}.

The \texttt{AuctionAgent} advances to slot $t+1$ only after all
\texttt{INFORM} messages have been received. This synchronous-round structure
removes asynchronous timing as a source of KPI differences. Because all four
strategies are evaluated on the same community and the same input profiles, the
remaining differences reflect the strategy rather than the scenario.

\subsection{Resource Models}
\label{subsec:devices}

Each prosumer owns a subset of four resource types. The net energy balance
determining surplus or deficit before storage dispatch is
\begin{equation}
E_\text{net}(t) = E_\text{PV}(t) - E_\text{base}(t) - E_\text{HP}(t),
\label{eq:balance}
\end{equation}
where $E_\text{PV}$ is \ac{pv} generation, $E_\text{base}$ is inflexible base
load, and $E_\text{HP}$ is \ac{hp} consumption.  \ac{bess} and
\ac{ev} dispatch are strategy-controlled.

\textit{Photovoltaics.} \Ac{pv} generation per interval follows
\begin{equation}
E_\text{PV}(t) = P_\text{peak} \cdot G(t) \cdot \Delta t,
\label{eq:pv}
\end{equation}
where $G(t) \in [0,1]$ is the normalized irradiance 
for Hamburg (2019)
\cite{PFE_LongTermpatternsofEuropeanPV_2016} and $P_\text{peak} = 5\,\text{kWp}$
per prosumer.

\textit{Base load and heat pump.} The H0 residential standard load profile
\cite{BDEW_StandardLastprofileStrom_2025} is applied, scaled to 3\,000\,kWh/a. The
\ac{hp} uses the WPE standard profile with the VDN TMZ block-heating method, a standardized German heat-pump load profile 
\cite{VDN_StepByStep}, scaled to 4\,000\,kWh/a. \Ac{hp} operation is treated
as non-flexible; demand-side management of \acp{hp} is addressed in \cite{MEIERS2025_Controlstrategies}.

\textit{Battery energy storage system.} The \ac{soc} evolves as
\begin{equation}
\text{SOC}(t) = \text{SOC}(t{-}1)
  + \frac{\eta_\text{ch}\,P_\text{ch}(t)\,\Delta t}{C_\text{bat}}
  - \frac{P_\text{dis}(t)\,\Delta t}{\eta_\text{dis}\,C_\text{bat}},
\label{eq:soc}
\end{equation}
with capacity $C_\text{bat} = 10\,\text{kWh}$, maximum power $P_\text{max} =
5\,\text{kW}$, efficiency $\eta_\text{ch} = \eta_\text{dis} = 0.95$, and
\ac{soc} bounds $[0.10,\,0.90]$ to protect cycle life.

\textit{Electric vehicle.} The same \ac{soc} model (\autoref{eq:soc}) applies with
$C_\text{bat} = 60\,\text{kWh}$ and $P_\text{max} = 11\,\text{kW}$.
Availability flag $a(t) \in \{0,1\}$ encodes a commuter pattern absent from
07:00 to 17:00. Daily energy demand is 11.7\,kWh (approximately 65\,km/day) \cite{Nobis2018_MobilitaetinDeutschland}.

\subsection{Market Clearing Mechanism}
\label{subsec:market}

Market clearing employs a uniform-price double-sided call auction once per
interval \cite{Gode1993_AllocativeEfficiencyofMarketswithZeroIntelligenceTraders}.
Sell orders are sorted by ascending price and buy orders by descending price
(merit order). The lists are matched pairwise from the top, each step trading the
smaller of the two remaining quantities and advancing past any filled order,
until the first pair with \mbox{$p_\text{ask} > p_\text{bid}$}. The marginal pair
is the last pair that traded, and the uniform clearing price is the midpoint of
its ask and bid,
\begin{equation}
p_\text{clear}(t) =
  \tfrac{1}{2}\bigl(p_\text{ask,marginal}(t) + p_\text{bid,marginal}(t)\bigr).
\label{eq:pclear}
\end{equation}

All matched volume in interval $t$ settles at $p_\text{clear}(t)$. A price
corridor $[p_\text{min},\,p_\text{max}] = [0.08,\,0.36]\,\text{EUR/kWh}$ bounds
all submitted prices, where $p_\text{min}$ equals the statutory feed-in tariff
for small \ac{pv} installations and $p_\text{max}$ the retail tariff for
residential customers \cite{Kirschen2004_FundamentalsOfPowerSystemEconomics}.
Unmatched sell volume is exported to the external grid at $p_\text{min}$;
unmatched buy volume is imported at $p_\text{max}$. Every \ac{lem} transaction
settled within the corridor produces a bilateral improvement relative to the
grid-tariff alternative, providing the fundamental incentive for participation.
The corridor reflects only the feed-in to retail spread. Taxes, network
charges, levies, and transaction costs are not modeled.
The market mechanism is implemented
as a centralized service within the \texttt{AuctionAgent} (see \autoref{subsec:arch}).

\subsection{Performance Metrics}
\label{subsec:kpis}

Nine \acp{kpi} characterize strategy outcomes:
\begin{itemize}[nosep, leftmargin=*, itemsep=1pt]
  \item {\Ac{ad}}: share of consumption covered locally without grid
    import \cite{Ortleb2022_OperationStrategiesSmartGridReady}.
  \item {\Ac{scr}}: fraction of \ac{pv} generation consumed locally
    \cite{Ortleb2022_OperationStrategiesSmartGridReady}.
  \item {\Ac{lem} quota}: share of total energy volume settled within the
    \ac{lem} \cite{Kilthau2023_MetricCooperativeCompetitiveCIRED}.
  \item {$C_\text{total}$}: aggregate net grid payments over the
    simulated day.
  \item {$p_\text{clear}$}: uniform settlement price per interval
    (\autoref{eq:pclear}).
  \item {\Ac{fgp}}: cost saving relative to a no-market counterfactual,
    in which all surplus \ac{pv} is exported at $p_\text{min}$ and all
    deficit is covered by grid import at $p_\text{max}$,
    summed over all community members
    \cite{Kilthau2023_MetricCooperativeCompetitiveCIRED}.
  \item {\Ac{mr}}: fraction of submitted bid volume cleared within the
    \ac{lem}.
  \item {\ac{bess} cycles}: storage utilization in equivalent full
    cycles per day.
  \item {\ac{lem} volume}: total energy quantity traded per simulated
    day.
\end{itemize}

\section{Market Strategies}
\label{sec:strategies}

The four strategies are defined along two dimensions: resource control and bid pricing.

\textit{ZIC (zero-intelligence constrained).}
Bid prices are drawn uniformly at random from the price corridor, following the
ZIC formulation of
\citet{Gode1993_AllocativeEfficiencyofMarketswithZeroIntelligenceTraders}.
Device control applies a basic cascade: \ac{pv} surplus charges the 
\ac{bess} up to its upper \ac{soc} bound. Remaining surplus is offered to the
market as a sell bid. On a deficit, the \ac{bess} discharges first. If the
\ac{bess} is exhausted or at its lower \ac{soc} bound, the agent submits a
buy bid to the market. If no match is found, the residual is settled with the
grid. ZIC provides a minimal-intelligence reference that isolates the
contribution of market structure from any strategic intent, and its
near-competitive efficiency in homogeneous populations (established by
\citet{Gode1993_AllocativeEfficiencyofMarketswithZeroIntelligenceTraders})
makes it a meaningful performance floor rather than a trivially weak baseline.

\textit{EVB (edge-value bidding).}
Sell bids are fixed at $p_\text{min}$ and buy bids at $p_\text{max}$. Every
submitted pair satisfies the matching condition, and clearing settles at
$p_\text{clear} = 0.22\,\text{EUR/kWh}$ per \autoref{eq:pclear}. This
guarantees full market participation but suppresses price discovery. Device
control is identical to ZIC, so EVB versus ZIC isolates the bid pricing
effect.

\textit{ESC (extended storage cascade).}
The resource control dimension is extended by adding a direct charge path from
surplus \ac{pv} to the \ac{ev} once the stationary \ac{bess} is full.
When surplus $\Delta E(t) > 0$ remains after \ac{bess} saturation and the
vehicle is available, the \ac{ev} receives
\begin{equation}
P_\text{ev,ch}(t) = \min\!\left(
  \tfrac{\Delta E(t)}{\Delta t},\;
  P_\text{ev,max} a(t),\;
  \tfrac{(\text{SOC}_\text{max} - \text{SOC}_\text{ev})\,C_\text{ev}}
       {\eta_\text{ch}\,\Delta t}
\right).
\label{eq:esn}
\end{equation}

Any $\Delta E(t)$ remaining after both storage paths are saturated is offered
to the market. Bid pricing follows the ZIP (zero-intelligence plus) principle
of \citet{Cliff1997_MinimalIntelligenceAgentsforBargaining}: each agent adjusts
its price stepwise by $\Delta p = 0.02\,\text{EUR/kWh}$ based on the previous
interval's clearing outcome. Sellers lower their ask when the previous clearing
price fell below their ask; buyers raise their bid when it exceeded their bid.
Ask prices thereby converge toward $p_\text{min}$ during high-supply intervals
and rise toward scarcity levels otherwise. ESC applies this mechanism uniformly
with a sell floor of $0.09\,\text{EUR/kWh}$ for all portfolio classes.
The EVB-to-ESC step therefore combines resource control and pricing changes;
the ESC-to-MA step isolates the portfolio-specific resource dispatch logic.

\textit{MA (market-adaptive strategy).}
Two portfolio-specific dispatch rules extend the resource control of ESC.
For prosumers (PV with BESS, PVB), BESS discharge for market
sale is inhibited when the most recent clearing price, observed in the previous
interval, falls below
\begin{equation}
p_\text{dis,min}
  = \frac{p_\text{min}}{\eta_\text{ch} \cdot \eta_\text{dis}}
  = \frac{0.08\,\text{EUR/kWh}}{0.95^{2}}
  \approx 0.089\,\text{EUR/kWh},
\label{eq:pdismin}
\end{equation}
under which selling stored energy yields less revenue than a direct grid feed-in
at $p_\text{min}$, producing a net loss over the charge-discharge round trip.
For PEB prosumers (\ac{pv} with both \ac{bess} and \ac{ev}), a dynamic
\ac{soc} reserve protects the anticipated remaining base load and \ac{hp}
demand until end-of-day, shrinking slot by slot as remaining demand decreases
and progressively freeing storage capacity for arbitrage.

Bid pricing uses the same ZIP mechanism as ESC, but with portfolio-differentiated
sell floors: $0.089\,\text{EUR/kWh}$ for prosumers with a \ac{bess}
and $p_\text{min} = 0.08\,\text{EUR/kWh}$ for those without.

\autoref{tab:strategies} summarizes the four strategies along their two dimensions.

\begin{table}[h]
\caption{Overview of the four market strategies across the resource control and
bid pricing dimensions.}
\label{tab:strategies}
\footnotesize
\setlength{\tabcolsep}{3pt}
\begin{tabularx}{\columnwidth}{lLL}
\toprule
Strategy & Device control & Price formation \\
\midrule
ZIC & Basic cascade (\ac{bess} priority) & Random in $[p_\text{min},\,p_\text{max}]$ \\
EVB & Basic cascade (\ac{bess} priority) & Boundary prices \\
ESC & Extended cascade (\ac{ev} priority + \ac{bess}$\to$\ac{ev} discharge) & Adaptive (ZIP-like) \\
MA  & Portfolio-specific reserve logic, price-controlled discharge & Adaptive (ZIP-like) \\
\bottomrule
\end{tabularx}
\end{table}

\section{Evaluation}
\label{sec:eval}

\subsection{Setup}
\label{subsec:setup}

The community consists of 33 prosumers distributed across an IEEE 33-bus network
topology.
Network
power flow constraints are not enforced. The topology serves as a spatial mapping
for prosumer assignment, and congestion management is addressed in prior work, such as \citet{Kil_IntegratingPeer-to-PeerEnergyTradingandFlexibility_2023}.
Portfolio assignment follows penetration rates drawn from
\citet{Kil_ADecentralizedOptimizationApproachForScalable_2025}: 67\,\% of
prosumers own \ac{pv}, 65\,\% own an \ac{ev}, and 80\,\% own an \ac{hp}.
The \ac{bess} rate of 50\,\% of \ac{pv} prosumers is an assumption of this
work, chosen to generate a diverse mix of classes. All classes include
an \ac{hp} and base load. \autoref{tab:agents} lists the five
resulting portfolio classes. All four strategies are evaluated independently on
the same community composition using Hamburg irradiance profiles.
The simulation code is publicly available.%
\footnote{\url{https://github.com/lukas-wagner/Market_Strategy_Evaluation}}

\begin{table}[h]
\caption{Prosumer portfolio classes with resource composition.}
\label{tab:agents}
\footnotesize
\setlength{\tabcolsep}{5pt}
\begin{tabularx}{\columnwidth}{lYYYY}
\toprule
Class    & Count & PV         & BESS       & EV         \\
\midrule
PEB      & 8     & \checkmark & \checkmark & \checkmark \\
PVB      & 3     & \checkmark & \checkmark & $-$        \\
PVEV     & 6     & \checkmark & $-$        & \checkmark \\
PVonly   & 5     & \checkmark & $-$        & $-$        \\
Consumer & 11    & $-$        & $-$        & $-$        \\
\bottomrule
\end{tabularx}
\end{table}

\subsection{Community-Level Strategy Comparison}
\label{subsec:results}

\begin{figure}[h]
\centering
\subfloat[Autarky (AD), self-consumption (SCR) \& LEM quota (LQ) \label{fig:kpis:autarky}]{%
\includegraphics{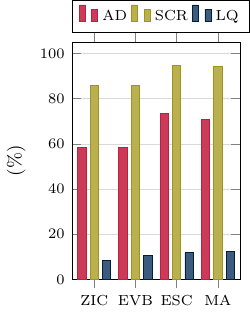}}%
\hfill
\subfloat[Cost \& financial gain (FGP) \label{fig:kpis:cost}]{%
\includegraphics{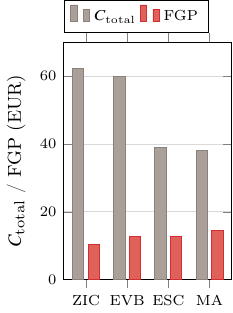}}
\caption{Community-level results for the four strategies.}
\label{fig:kpis}
\end{figure}

\autoref{fig:kpis} summarizes the primary community-level results
(metric definitions in \autoref{subsec:kpis}).

\textit{Autarky, self-consumption, and \ac{lem} quota (\autoref{fig:kpis:autarky}).}
ZIC and EVB yield identical autarky degree (58.5\,\%) and self-consumption rate (85.8\,\%),
confirming that bid pricing alone does not affect physical energy coverage.
ESC raises both metrics substantially (autarky to 73.6\,\%, self-consumption
to 94.8\,\%) by routing surplus \ac{pv} into \ac{ev} storage before
offering the remainder to the market. MA follows at 70.8\,\% and 94.3\,\%:
the dynamic \ac{soc} reserve trades a modest autarky penalty for
price-driven dispatch flexibility. The \ac{lem} quota rises from 8.6\,\% (ZIC) to
12.6\,\% (MA), with the step from ZIC to EVB (8.6 to 10.7\,\%) reflecting
guaranteed bid matching, and the further increase under ESC and MA driven
by adaptive pricing expanding the set of cleared pairs.

\textit{Cost and financial gain (\autoref{fig:kpis:cost}).}
Regarding RQ1, pricing strategy has a modest isolated effect: switching from
random to deterministic boundary prices (ZIC to EVB) reduces $C_\text{total}$
by only $2.46\,\text{EUR}$ (3.9\,\%), confirming that price anchoring alone
yields limited cost reduction. The EVB-to-ESC step combines the extended storage
cascade with adaptive pricing and achieves the single largest improvement:
$C_\text{total}$ drops to $39.06\,\text{EUR}$, a combined reduction of
$20.86\,\text{EUR}$ (34.8\,\%). Because this step changes both the storage
cascade and the pricing rule, it quantifies the combined ESC design, not the
extended cascade in isolation. Only the ZIC-to-EVB step isolates a single
dimension, pricing. Switching from the uniform ESC cascade to the
portfolio-specific resource dispatch logic of MA provides a further $0.85\,\text{EUR}$
(2.2\,\%) saving. Combined, the full upgrade from ZIC to MA reduces community
expenditure by $24.17\,\text{EUR}$ (38.8\,\%) relative to ZIC. Community financial gain
through \ac{lem} participation increases in parallel: \ac{fgp} rises from
$10.28\,\text{EUR}$ (ZIC) to $14.40\,\text{EUR}$ (MA), a total improvement of
$4.12\,\text{EUR}$ (40.1\,\%). Storage utilization, market volume, clearing
price, and match rate are discussed in the following subsections.

\subsection{Storage Utilization, Market Volume, and Portfolio Effects}
\label{subsec:portfolio}

\begin{figure}[h]
\centering
\subfloat[BESS utilisation\label{fig:storage:bess}]{%
\includegraphics{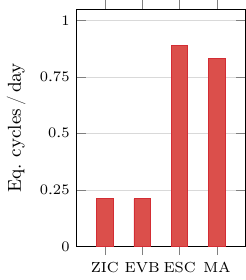}}%
\hfill
\subfloat[LEM volume\label{fig:storage:lem}]{%
\includegraphics{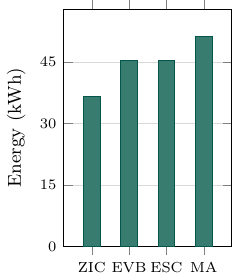}}
\caption{\ac{bess} equivalent full cycles per day (left) and total
energy volume settled within the \ac{lem} (right).}
\label{fig:storage}
\end{figure}

\autoref{fig:storage} shows battery utilization (\autoref{fig:storage:bess}),
the primary driver of the cost reductions in \autoref{fig:kpis:cost}, and
total traded market volume (\autoref{fig:storage:lem}), which reflects the
degree to which strategies activate the \ac{lem} as an exchange channel.

RQ2 asks how the \ac{ev} battery integration affects autarky across portfolio
classes. The answer is class-specific: the largest benefit accrues to
prosumers that possess both a \ac{bess} and an \ac{ev} (PEB class).
\autoref{tab:portfolio_ad} shows that PEB autarky rises from 50.6\,\% under
ZIC and EVB to 92.2\,\% under ESC, a gain of 41.6 percentage points. The
mechanism is the fourfold increase in \ac{bess} equivalent full cycles
(\autoref{fig:storage:bess}, 0.21 to 0.89 per day): by routing surplus \ac{pv}
into the \ac{ev} battery once the \ac{bess} is full (\autoref{eq:esn}), and
subsequently discharging the \ac{bess} into the \ac{ev} when the \ac{soc}
reserve permits, ESC fills both the 10\,kWh stationary and the 60\,kWh mobile
storage, covering evening demand without grid import. Prosumers with only a
\ac{pv} and an \ac{ev} (PVEV) see a moderate improvement, reaching
62--71\,\% autarky, since the \ac{ev} battery alone provides less intra-day
flexibility without a stationary buffer. Prosumers without an \ac{ev} (PVB)
gain no benefit from the cascade extension, as the \ac{ev} charge path does
not apply. Prosumers with only \ac{pv} remain near 45\,\% regardless of
strategy, with no storage to shift surplus across time.

\begin{table}[t]
\caption{AD of the PEB portfolio class vs.\ the community
average for all four strategies (summer day). Differences in percentage points.}
\label{tab:portfolio_ad}
\footnotesize
\setlength{\tabcolsep}{4pt}
\begin{tabularx}{\columnwidth}{lYYYY}
\toprule
                           & ZIC    & EVB    & ESC    & MA     \\
\midrule
PEB class AD (\%)          & 50.6   & 50.6   & 92.2   & 84.4   \\
Community average AD (\%)  & 58.5   & 58.5   & 73.6   & 70.8   \\
\midrule
Difference (pp)            & $-7.9$ & $-7.9$ & $+18.6$& $+13.6$\\
\bottomrule
\end{tabularx}
\end{table}

MA reaches only 84.4\,\% autarky for PEB, 7.9 percentage points below ESC:
the dynamic reserve logic (\autoref{eq:pdismin}) retains \ac{bess} capacity
for evening baseload coverage, preventing full \ac{ev} charging. The gap between PEB and the
community average grows from $-7.9$ percentage points under ZIC to $+18.6$
points under ESC, confirming that portfolio class and strategy are jointly
determining factors that cannot be assessed independently.

Despite the higher storage utilization, market volume is identical for EVB
and ESC at 45.5\,kWh (\autoref{fig:storage:lem}): the extended cascade
absorbs more energy locally rather than offering it to the market. MA reaches
the highest traded volume of 51.4\,kWh, as adaptive pricing attracts
additional matching bids from both sides.

All portfolio classes retain a positive \ac{fgp} under every strategy: market
participation always improves on the no-market counterfactual. However, the
distribution of gains shifts under MA. All \ac{pv}-owning classes receive
lower \ac{fgp} under MA than under EVB, because the lower clearing price
reduces per-unit sell-side revenue: PVonly by $0.23\,\text{EUR}$, PVEV by
$0.12\,\text{EUR}$, PEB by $0.09\,\text{EUR}$, and PVB by $0.07\,\text{EUR}$
per prosumer. Pure consumers are the sole net beneficiaries of adaptive
pricing, gaining an additional $0.41\,\text{EUR}$ per prosumer under MA
relative to EVB. The portfolio-level \ac{fgp} spread widens from
$0.47\,\text{EUR}$ under EVB to $0.97\,\text{EUR}$ under MA, indicating that
higher aggregate community gain comes at the cost of increased distributional
inequality. Under adaptive pricing, sellers accept lower ask prices to secure
a match, which compresses their per-unit revenue while buyers clear at a more
favorable price; the pricing rule therefore transfers value from prosumers with
generation assets to pure consumers rather than distributing gains uniformly.

\subsection{Clearing Price Dynamics and Value Distribution}
\label{subsec:dynamics}

\begin{figure}[h]
\centering
\subfloat[Avg.\ clearing price\label{fig:price:cp}]{%
\includegraphics{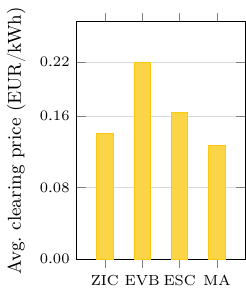}}%
\hfill
\subfloat[Match rate\label{fig:price:mr}]{%
\includegraphics{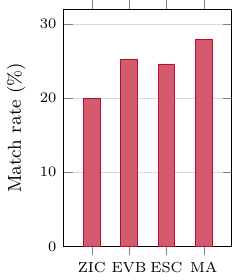}}
\caption{Volume-weighted average clearing price (left) and match rate (right)
for the four strategies.}
\label{fig:price}
\end{figure}

\begin{figure}[t]
\centering
\includegraphics{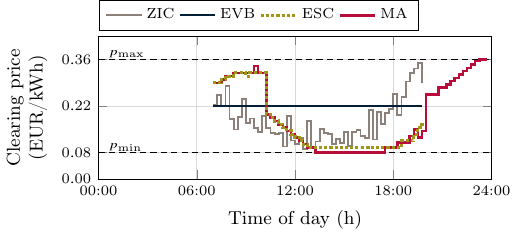}
\caption{Intraday clearing price per 15-min slot; only slots with
at least one matched trade are shown (gaps indicate no clearing).
ESC (dotted) and MA (solid) are plotted with overlapping styles to
keep both visible where prices coincide.
}
\label{fig:price:ts}
\end{figure}

\autoref{fig:price} summarizes the aggregate pricing outcomes; the intraday
dynamics are shown in \autoref{fig:price:ts}.

EVB serves as the deterministic reference: boundary prices fix the clearing
price at $0.22\,\text{EUR/kWh}$ regardless of supply-demand balance.
ZIC already undercuts this level ($0.14\,\text{EUR/kWh}$, \autoref{fig:price:cp}),
but not through strategic intent: random bids occasionally fall well below
the competitive level, matching at prices unfavorable for sellers. This is
the allocatively efficient but individually suboptimal outcome identified by
\citet{Gode1993_AllocativeEfficiencyofMarketswithZeroIntelligenceTraders}.
Adaptive pricing (ESC and MA) achieves lower prices through deliberate
convergence: ESC reaches $0.164\,\text{EUR/kWh}$ and MA $0.127\,\text{EUR/kWh}$,
a reduction of 42.3\,\% relative to EVB. ESC's match rate ($24.6\,\%$) falls
marginally below EVB ($25.2\,\%$): the extended storage cascade absorbs more
surplus locally, reducing the volume of sell bids submitted, while the adaptive
pricing mechanism requires several intervals to converge. During this phase,
ask prices may temporarily exceed buy bids and prevent clearing.
MA reverses this effect and raises the match rate to $27.9\,\%$
(\autoref{fig:price:mr}): once prices have converged, the lower ask floor
attracts buy bids that could not match at the EVB boundary, expanding the set of
feasible bid pairs.

\autoref{fig:price:ts} reveals the intraday mechanism behind the aggregate
numbers. ZIC produces a scattered price profile across the day: random sell bids
occasionally fall well below what a rational seller would accept, yielding
clearing prices that are irregular and often low
(volume-weighted average $0.14\,\text{EUR/kWh}$).
EVB is flat at $0.22\,\text{EUR/kWh}$ across all active slots: the
boundary-priced bids always resolve to the midpoint of the corridor,
independent of supply-demand balance. ESC and MA share an identical morning
profile ($0.29$--$0.32\,\text{EUR/kWh}$, 07:00--10:00) when \ac{pv}
generation is ramping and seller supply is thin. As irradiance peaks, both
strategies adapt ask prices downward: ESC stabilizes near
$0.095\,\text{EUR/kWh}$ (the round-trip efficiency sell floor), while MA
converges to $p_\text{min} = 0.08\,\text{EUR/kWh}$ for non-\ac{bess} sellers
whose floor is lower. The critical difference is trading duration. MA remains
active until 23:45 (64 slots vs.\ 51 for ESC), because price-controlled
\ac{bess} discharge in PVB prosumers generates sell volume in post-sunset hours.
Sellers in these slots face demand-side scarcity and can command prices up to
$p_\text{max}$; buyers simultaneously avoid the grid tariff. This evening
extension arises from MA's portfolio-specific resource dispatch logic for PVB prosumers,
which enables price-controlled \ac{bess} discharge that ESC's uniform cascade
does not provide, and explains the $1.66\,\text{EUR}$ \ac{fgp} advantage of
MA over ESC ($14.40$ vs.\ $12.74\,\text{EUR}$, \autoref{fig:kpis:cost}).

The clearing price and the traded volume act through two separate channels. At a fixed volume the clearing price only splits the local-trading savings between buyers and sellers, while the strategies also change the match rate and LEM volume, and thus how much saving exists to split. When a kilowatt-hour is traded locally, the buyer pays less than the grid retail tariff and the seller receives more than the feed-in tariff; however, a lower clearing price increases the buyer's saving at the expense of the seller's margin, so these effects offset at community level. Under MA, lower clearing prices shift $0.41\,\text{EUR}$ per prosumer from sellers to buyers compared to EVB, consistent with \citet{Nicolaisen2001_Marketpowerandefficiencyinaelectricitymarket}. Despite this redistribution, the community's total \ac{fgp} under MA rises by $1.67\,\text{EUR}$ compared to EVB, because more energy is traded locally rather than settled at regulated grid tariffs.

\subsection{Seasonal Dependency}
\label{subsec:seasonal}

The summer evaluation in \autoref{subsec:results} represents conditions favorable
to \ac{lem} activity: high \ac{pv} generation creates ample surplus for trading,
and the community contains both net sellers and net buyers in most intervals.
A winter evaluation (Hamburg, January) reveals a qualitatively different regime.
With irradiance at roughly one-sixth of summer peak values, \ac{pv} generation
is largely consumed immediately at the generating prosumer, and the self-consumption
rate reaches 95.1\,\% under all strategies. Surplus available for \ac{lem}
trading is minimal: \ac{lem} quota drops to 1.4\,\% and community cost is identical
for EVB and MA at $367.05\,\text{EUR}$, confirming that the pricing mechanism
is irrelevant when there is no surplus to trade.
Under MA, the clearing price rises to $0.35\,\text{EUR/kWh}$, approaching
$p_\text{max}$, because adaptive buyers compete for the scarce \ac{pv} supply
rather than sellers competing to sell. The \ac{lem} becomes effectively inactive
in winter, not because the mechanism fails, but because the physical
supply-demand balance produces no meaningful surplus: every kilowatt-hour of
\ac{pv} generation is already cheaper than a grid alternative and prosumers
consume it locally regardless of market signals.

Spring conditions (Hamburg, April) represent an intermediate regime. Moderate
\ac{pv} generation combined with residual space-heating demand creates a
supply-demand balance that is, in some respects, more favorable for intra-community
trading than peak summer: sufficient surplus exists to generate competitive sell
bids, while elevated consumption from heating loads maintains buy-side demand
at a level that absorbs most of the available surplus. Under EVB, \ac{lem} quota
reaches 17.4\,\% in April, notably above the 10.7\,\% observed in July. This
counter-intuitive result reflects that peak summer \ac{pv} surpluses can exceed
the \ac{lem}'s absorption capacity when all generators simultaneously produce
more than the community can consume, causing excess export to the grid even with
active market participation. A key reversal occurs for community \ac{fgp}: in April,
EVB achieves $23.59\,\text{EUR}$ while MA achieves only $22.45\,\text{EUR}$
($-1.14\,\text{EUR}$), the opposite of the summer relationship. With moderate
\ac{pv} penetration, sell-side competition is weaker, and sellers have no
incentive to adapt their ask prices downward; adaptive pricing thus fails to
expand the cleared volume sufficiently to offset the lower per-unit revenue.
These seasonal patterns imply that the optimal strategy for a prosumer varies
across seasons: boundary prices are better suited to spring conditions while
adaptive pricing delivers greater advantage in summer when supply-side
competition among sellers is most intense.

\section{Conclusion}
\label{sec:conclusion}

Prosumers participating in \acp{lem} face two intertwined decisions: how to dispatch flexible resources locally and at what price to bid into the market. Evaluating these dimensions jointly and in isolation is essential for understanding their respective contributions to community cost and financial gain.

This work presented an agent-based \ac{lem} simulation with 33 heterogeneous
prosumer agents participating in a uniform-price call auction under four
strategies of increasing complexity, evaluated across summer, winter, and
spring conditions. Resource control proved to be the primary lever: the extended
storage cascade reduced community cost by 37.4\,\% relative to the
zero-intelligence baseline, while market-adaptive pricing raised community financial gain
through \ac{lem} participation by 40.1\,\%. Autarky improvements were confined
to prosumers combining \ac{pv} with \acp{bess} and \acp{ev},
demonstrating that strategy effectiveness is inseparable from portfolio
composition. Adaptive pricing redistributes value from net sellers to net
buyers through lower clearing prices, whereas boundary pricing offers composition-independent
stability at the cost of lower aggregate gain. Seasonal analysis showed that
this advantage is not universal: in spring, weaker sell-side competition renders
adaptive price adaptation ineffective, and boundary pricing achieves higher community
financial gain. In winter, minimal \ac{pv} surplus makes the pricing mechanism
largely irrelevant, and strategy differences vanish. The optimal choice therefore
depends on the season, shown directly, and on community composition to the
extent indicated by the portfolio-class differences, motivating portfolio- and
season-aware strategy selection, with a composition sensitivity analysis, as
a direction for future work.

Several limitations apply. The simulation assumes perfect information with no forecast uncertainty, identical \ac{pv} sizes across prosumers, deterministic \ac{ev} availability, and a single fixed community composition, so the community-composition dependence is shown only through portfolio-class differences within that community. All agents use the same strategy per run, so the results characterize homogeneous strategy deployment, not the equilibrium under independently chosen strategies. Mixed-strategy scenarios are not evaluated. ZIC draws random bids, so its reported results are a single realization rather than a mean over repeated draws, and it is the only stochastic strategy. \ac{bess} degradation costs are not modeled: the fourfold increase in cycling under the extended cascade (0.21 to 0.89 equivalent full cycles per day, roughly 77 to 325 cycles per year) would narrow its economic advantage once aging is accounted for. Network constraints are not enforced.

Future work will pursue field deployment to validate the simulation results under real forecast uncertainty and dynamic tariff structures. Investigating heterogeneous strategy populations would reveal competitive dynamics and equilibrium properties of the market. Additionally, the observed portfolio and seasonal dependency motivates the development of adaptive mechanisms that select or parameterize strategies based on community composition and prevailing supply conditions.

\bibliographystyle{IEEEtranN}
\scriptsize{\bibliography{literature}}

\end{document}